\documentclass[9pt,twocolumn,twoside]{opticajnl}
\journal{opticajournal} % use for journal or Optica Open submissions

\setboolean{shortarticle}{true}
\usepackage{lineno}
\title{Symmetry-Broken Cavity Solitons and Collective Polarization Conformity in Fabry–Pérot Kerr Resonators}
\author[1]{Yohann G. Sanvert}
\author[2,3,4]{Abdullah Alabbadi}
\author[2]{Lewis Hill}
\author[5]{Yuandi Xu}
\author[5]{Gang Xu}
\author[6]{Gian-Luca Oppo}
\author[7]{Stéphane Coen}
\author[1]{Erwan Lucas}
\author[2,3]{Pascal Del’Haye}
\author[1,*]{Julien Fatome}

\affil[1]{Université Bourgogne Europe, CNRS, Laboratoire Interdisciplinaire Carnot de Bourgogne ICB UMR 6303, 21000 Dijon, France}
\affil[2]{Max Planck Institute for the Science of Light, Staudtstr. 2, 91058 Erlangen, Germany}
\affil[3]{Department of Physics, Friedrich Alexander University Erlangen-Nuremberg, 91058 Erlangen, Germany}
\affil[4]{Department of Physics, Faculty of Science, Alexandria University, Moharram Bek, 21511 Alexandria, Egypt}
\affil[5]{School of Optical and Electronic information, Huazhong University of Science and Technology, Wuhan, China}
\affil[6]{SUPA and Department of Physics, University of Strathclyde, Glasgow G4 0NG, Scotland, UK}
\affil[7]{Physics Department, The University of Auckland, Private Bag 92019, Auckland 1142, New Zealand}

\affil[*]{Julien.Fatome@u-bourgogne.fr}

\begin{abstract}
We report on the experimental generation of polarization symmetry-broken cavity solitons (CSs) in a passive, fiber-based, coherently-driven, Fabry–Pérot (FP) Kerr resonator. Polarization-resolved measurements reveal the spontaneous transition of initially symmetric CSs into asymmetrical vectorial states, triggered by a cross-phase modulation-induced polarization bifurcation. Most notably, due to counter-propagation of light occurring in FP resonators, we unveil a collective polarization conformity effect, whereby multiple CSs circulating in the cavity converge to the same asymmetric polarization state once their number exceeds a certain threshold. These results demonstrate that Fabry–Pérot resonators support novel collective soliton dynamics that are absent in ring architectures.
\end{abstract}

\setboolean{displaycopyright}{false} % Do not include copyright or licensing information in submission.

\begin{document}

\maketitle

%\section{Introduction}
{Temporal cavity solitons (CSs) are self-localized dissipative structures that can persist indefinitely in coherently-driven, nonlinear passive Kerr resonators \cite{Leo2010,Herr2013}. They arise from a dual balance between chromatic dispersion and nonlinearity on the one hand, and loss and gain on the other \cite{Wabnitz1993, oppo_theory_2024}. CSs are well known to underpin the formation of broadband optical frequency combs (OFCs) in microresonators and have attracted significant attention over the past decade due to their promising potential in a plethora of applications, including high-capacity optical communications, laser ranging, and spectroscopy \cite{kippenberg_dissipative_2018,pasquazi_micro-combs_2018}. CSs have been predominantly studied as single-component (scalar) structures, and have been found to be ubiquitous in a wide variety of platforms, including passive and active fiber cavities, integrated micro-rings, Fabry–Pérot (FP) resonators, as well as semiconductor-based and free-space cavities \cite{Leo2010,Herr2013,Wabnitz1993, oppo_theory_2024, kippenberg_dissipative_2018,pasquazi_micro-combs_2018,Englebert2020,Obrzud2017,pedaci_positioning_2006,lilienfein_temporal_2019}. It was only recently that two-component temporal CSs have been reported in fiber cavities by leveraging the vectorial nature of light \cite{xu_spontaneous_2021,Hill2020, yang_polarization_2025}. More specifically, it has been shown that cross-polarized nonlinear coupling in Kerr ring resonators can give rise to spontaneous symmetry-breaking (SSB) of polarization, enabling the formation of interlocked, polarized structures \cite{xu_spontaneous_2021,lucas_polarization_2025, Coen_fat_2024, hill2020effects,campbell2024frequency}.
Here, we report the experimental generation of polarization symmetry-broken cavity solitons (SB-CSs) in a Fabry–Pérot (FP) Kerr resonator consisting of a few-meter-long anomalous dispersive optical fiber encapsulated between two dielectric Bragg mirrors.
This FP architecture provides an ideal resonator platform, combining the flexibility, stability, and inter-connectivity of fiber-based systems with free spectral ranges (FSRs) spanning from hundreds of MHz to tens of GHz, while preserving high quality factors \cite{Obrzud2017, yang_polarization_2025,jia_photonic_2020, musgrave_microcombs_2023,li_ultrashort_2024,bunel_28_2024,moroney_kerr_2022}. 
The SB-CSs are generated under a recently discovered regime of topological protection of the SSB phenomenon, which imparts strong robustness against external perturbations \cite{Coen_fat_2024}. Moreover, we reveal for the first time a distinctive collective behavior --- polarization conformity --- among CSs, which, in contrast to ring architectures, arises specifically from the counter-propagating nature of light in FP resonators \cite{hill_symmetry_2024}.

%\section{Modeling}
 The system under study can be modeled by two coupled Lugiato-Lefever-like Equations (LLEs) describing the slow-time evolution of both left- and right-circular polarization components of the circulating field $E_\pm$~\cite{cole_theory_2018,hill_symmetry_2024}.
 \begin{multline}
    \dfrac{\partial E_{\pm}}{\partial t} = \sqrt{\frac{X}{2}}
        -E_{\pm} - i\Delta E_{\pm} - i\dfrac{\partial^2 E_{\pm}}{\partial \tau^2} + i(|E_{\pm}|^2 + B |E_{\mp}|^2)E_{\pm} \\
               + i[(2 \langle|E_{\pm}|^2\rangle + B \langle|E_{\mp}|^2\rangle) E_{\pm} + B \langle E_{\pm}E_{\mp}^{*}\rangle E_{\mp}]  \,.
    \label{eq:LLEs}
\end{multline}
$t$ and $\tau$ are respectively the slow and fast time coordinates, $X$ is the total driving power, $\Delta$ the cavity detuning, and $B$ the cross-polarization coefficient \cite{hill2020effects}, evaluated at about 1.2 in the present experiments. Note that in contrast to ring architectures, the equations above include three extra  nonlocal nonlinear terms, which stem from the counter-propagating nature of FP resonators, with angled brackets denoting temporal averages over a full resonator roundtrip~\cite{cole_theory_2018,hill_symmetry_2024}.

To investigate the emergence of SB-CSs in FP Kerr resonators, we first compute the steady-state solutions of Eqs.~(\ref{eq:LLEs}) using a continuation method, as shown in Fig.~\ref{fig:bifurcation}(a). Unlike ring resonators, the bifurcation curves of SB-CSs in FPs depend explicitly on the filling factor, \textit{i.e.} the number of CSs circulating in the cavity. This behavior originates from the nonlocal terms of Eqs.~(\ref{eq:LLEs}), which introduce an additional effective detuning range of the nonlinear resonance and enhances the contrast between the two CS components, as the number of CSs increases~\cite{cole_theory_2018,hill_symmetry_2024}. For example, a single SB-CS can exist up to $\Delta = 12.3$, whereas 20 SB-CSs remain stable up to $\Delta = 16.1$. Panels~(b--c) further display the corresponding temporal intensity profiles $I_\pm = |E_\pm|^2$. Note that to observe such nonlinear dynamics, regular spacings between CSs is not mandatory.
\begin{figure}[h]
\centering
\includegraphics[width=\linewidth]{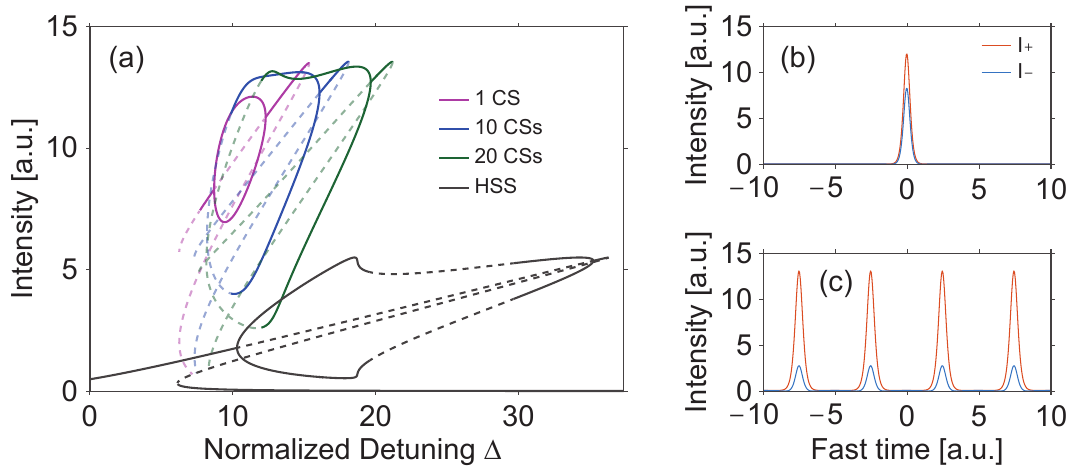}
\caption{\justifying{(a) Bifurcation diagrams computed from Eqs.~(\ref{eq:LLEs}), showing the intensity of homogeneous steady states (HSS) compared to SB-CSs for 1, 10, and 20 CSs circulating in the FP resonator. Unstable or non-stationary solutions are indicated by dashed lines. Parameters are $X = 11$ and $B = 1.2$. Panels~(b--c) compare the intensity profiles of SB-CSs for a single CS ($\Delta = 10$) and for 20 CSs ($\Delta = 12.5$), respectively.}}
\label{fig:bifurcation}
\end{figure}

%\section{Experimental setup}
{Figure~\ref{fig:setup} displays the experimental setup. The FP resonator consists of a 5-m-long segment of single-mode fiber [free-spectral-range (FSR) of 20~MHz],  characterized by an anomalous dispersion coefficient $\beta_2 = -21~\mathrm{ps^2/km}$ and a Kerr coefficient $\gamma = 1.3~\mathrm{/W/km}$, butt-coupled to FC/PC connectors whose facets are coated with multilayer dielectric mirrors \cite{moroney_kerr_2022}. These mirrors were fabricated through successive deposition of thin films, alternating tantalum pentoxide ($\mathrm{Ta_{2}O_{5}}$), high refractive index) and silicon dioxide (low refractive index) layers. Particular attention was given to tailoring the mirror reflectivity ($99.2\%$) to obtain a high cavity finesse of 153 (resonance linewidth of 154 kHz, \textit{cf.} inset in Fig.~\ref{fig:NLresonance}).
The FP resonator also incorporates an intra-cavity polarization controller positioned on one side of the cavity, introducing a $\pi/2$ phase shift between the two orthogonal polarization eigenmodes. This results in an overall $\pi$-phase shift per roundtrip (RT) due to the back-and-forth propagation in the FP. This configuration places the system within the recently discovered regime of topological protection of the SSB phenomenon \cite{Coen_fat_2024}, for which the presence of this localized phase defect makes the light’s handedness to swap at every cavity RT, leading to a self-symmetrization of the nonlinear dynamics on a two-RT cycle.
The FP resonator is coherently driven using ns-duration bursts of 10-ps optical pulses temporally gated (with an intensity modulator, IM) out of a 20-GHz repetition rate source \cite{fatome_20ghz_2006}, synchronized to a multiple of the cavity's FSR. This pulsed driving scheme allows for high peak power injection, while providing precise control over the number of CSs that can be generated in the cavity --- an essential feature for investigating collective behaviors. Moreover, a phase modulator (PM1), driven by an arbitrary waveform generator (AWG), enables us to perturb the driving beam so as to effectively write or erase sequences of CSs within the FP \cite{jang_writing_2015}. In addition, the cavity detuning $\Delta$ was adjusted and actively stabilized using a sinusoidal phase modulation scheme (PM2) applied to the driving beam, with the induced spectral sidebands serving as Pound–Drever–Hall (PDH) signal to lock the driving laser frequency on a cavity resonance (see the gray curve in Fig.~\ref{fig:NLresonance}). Finally, stability of the experimental setup was ensured by thermally and acoustically isolating the FP resonator within a dedicated enclosure.}

\begin{figure}[h]
\centering
\includegraphics[width=\linewidth]{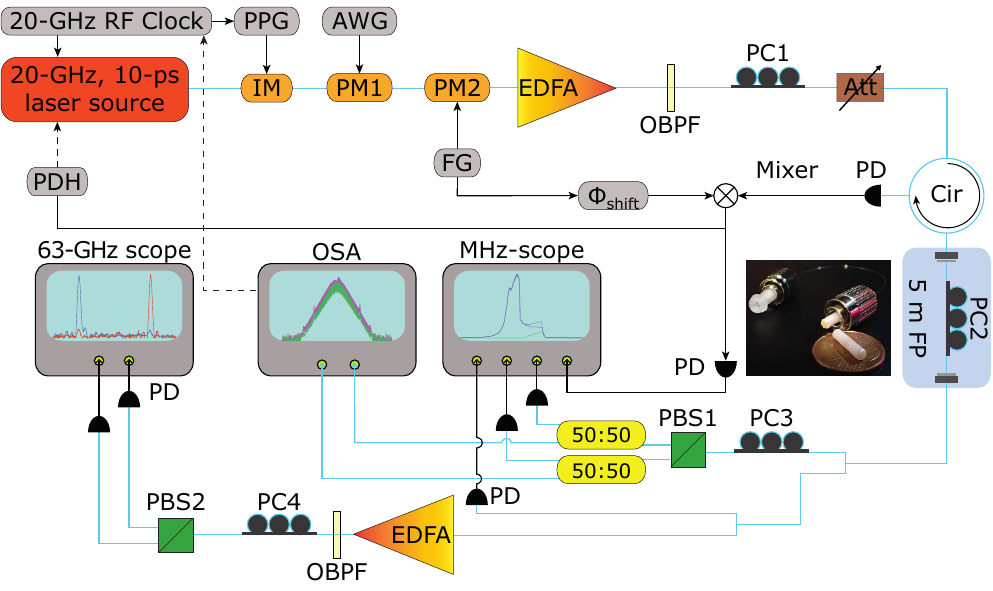}
\caption{\justifying{Experimental setup. PPG: Pulse pattern generator, IM: Intensity modulator, PM: Phase modulators, AWG: arbitrary waveform generator, FG: Function generator, EDFA: Erbium doped fiber amplifiers, OBPF: Optical band-pass filter, PC: Polarization controllers, Att: Variable attenuator, Cir: Optical circulator, PBS: Polarizing beam splitters, PD: Photo-detectors, OSA: Optical spectrum analyzer, PDH: Pound–Drever–Hall. The inset shows a photo of the FP resonator mirrors.}}
\label{fig:setup}
\end{figure}
%\section{ Experimental results}
Figure~\ref{fig:NLresonance} shows the nonlinear resonance of the FP cavity obtained under injection of a 5-ns temporal burst containing 200 pulses, with a driving peak power of 4~W ($X\sim32$)}. The FP is driven along one of its principal polarization axes (the $x$-component). The total average intra-cavity power (blue curve) exhibits a characteristic \textit{soliton step} for detuning values above 10, a well-known signature of CSs formation \cite{Herr2013}. More importantly, the occurrence of SSB can be inferred from the behavior of the polarization-resolved components of the resonance. Specifically, beyond a detuning threshold $\Delta \sim 18$, the $x$-polarized component (purple curve) saturates, while a significant $y$-polarized component (green curve) emerges across the soliton step --- despite being undriven. This observation, well confirmed by numerical simulations depicted with pink solid lines, unveils that polarization SSB of CSs is associated with the parametric generation of an orthogonal, background-free, pulse component.

\begin{figure}[h!]
\centering
\includegraphics[width=\linewidth]{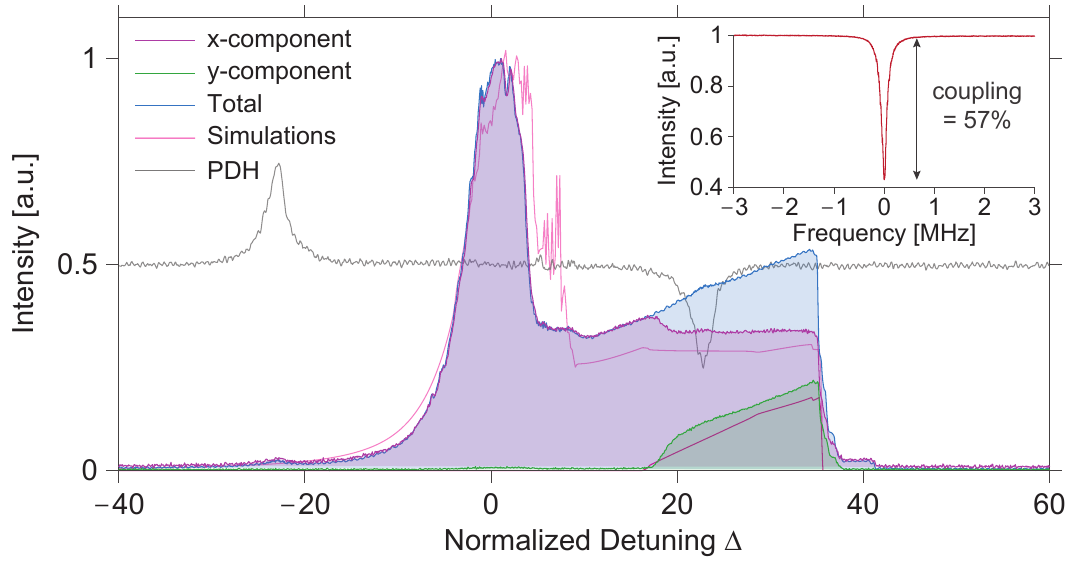}
\caption{\justifying{Nonlinear resonance measured for a driving peak power of 4~W. The purple curve shows the resonance along the driving polarization state ($x$), while the green curve displays the orthogonal (undriven) $y$-component. The blue curve is the total power, and the pink lines correspond to numerical simulations. The gray curve indicates the PDH signal used to lock the cavity at a specific detuning. The inset depicts the linear reflected response of the FP resonator, revealing that on resonance, up to $57\%$ of the input power is coupled into the cavity.}}
\label{fig:NLresonance}
\end{figure}

\begin{figure}[h!]
\centering
\includegraphics[width=\linewidth]{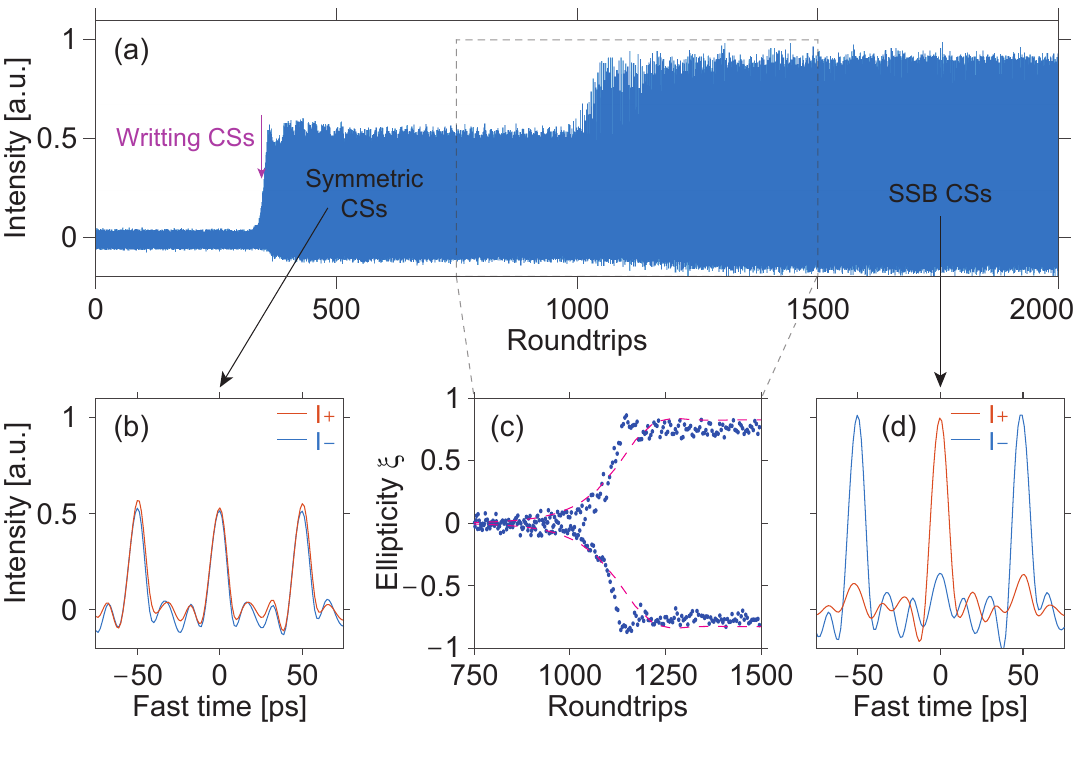}
\caption{\justifying{(a) Evolution of a sequence of CSs over successive roundtrips, showing the writing of symmetric CSs followed by their bifurcation into symmetry-broken states (only the $I_-$ polarization component is represented for clarity purpose).  (b, d) Temporal intensity profiles of symmetric and SB-CSs, respectively. (c) Ellipticity of the two first CSs shown in Panel (b) and (d) around the bifurcation point as highlighted by the dashed box in Panel (a) [dark blue dots: experimental data; pink dashed line: numerical simulations].}}
\label{fig:asym_sym}
\end{figure}

%\subsection{Symmetric-to-Asymmetric transition.}
We then investigate the excitation and transition of initially symmetric CSs into asymmetric vectorial solitons. To this end, the cavity detuning was locked along the soliton step near $\Delta~=~23$ (see the PDH signal in Fig.~\ref{fig:NLresonance}), and CSs were excited in the resonator via phase perturbations applied on the driving beam. The slow-time evolution of the intra-cavity polarization component $I_-$ is reported in Fig.~\ref{fig:asym_sym}a, ($I_+$ is omitted for convenience).
The phase perturbation is applied after about 300 RTs. CSs are successfully written in the cavity and initially appear in a symmetric state, as confirmed by equal intensities $I_+ = I_-$ in the temporal intensity profiles shown in Panel (b). Note that the temporal width of CSs is restricted to 13 ps due to the bandwidth limitation (63 GHz) of our oscilloscope. These stationary symmetric CSs propagate for about 700 RTs before undergoing a spontaneous bifurcation toward a fully asymmetric state around RT~$\#$1000, as unambiguously confirmed by the final temporal profiles shown in Panel (d). This transition is also confirmed by the sudden elevation of peak power in the temporal trace of $I_-$ in Panel (a) --- a clear signature of energy flow toward a dominant polarization component.
To further analyze this transition, Panel (c) presents the evolution of the polarization ellipticity $\xi = (I_+ - I_-)/(I_+ + I_-)$ of the two first CSs shown in (b) over successive RTs near the bifurcation point. Starting from a scalar, symmetric state ($\xi \approx 0$), the CSs abruptly break their polarization symmetry and evolve towards nearly  circularly polarized states ($\xi \approx \pm 1$). Note that this bifurcation is triggered by quantum noise, which randomly determines the final handedness of SB-CSs.

\begin{figure}[h]
\centering
\includegraphics[width=\linewidth]{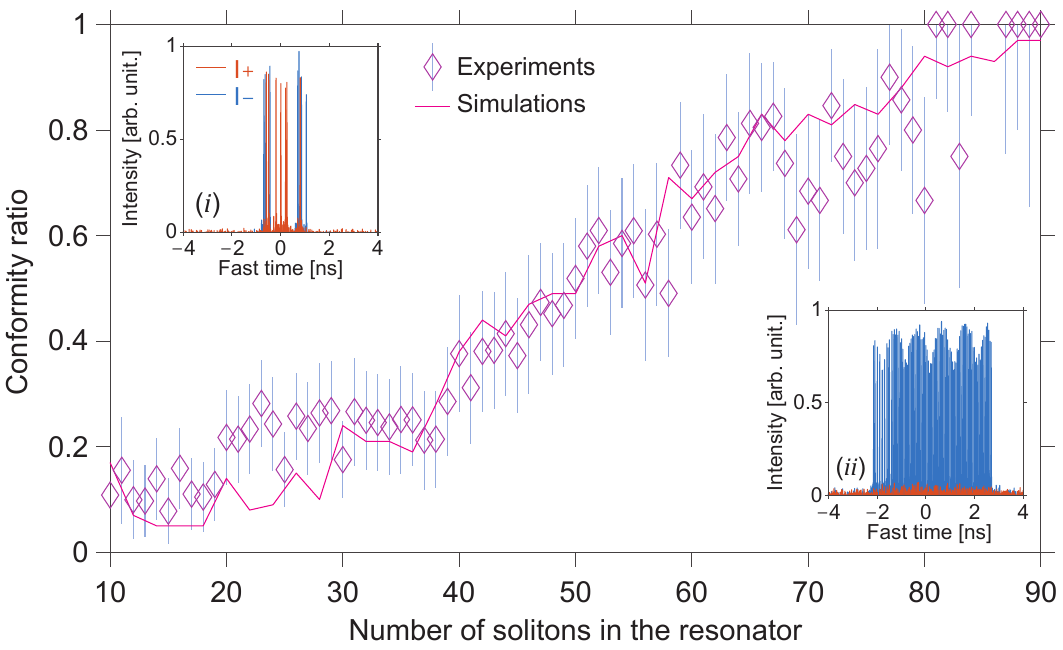}
\caption{\justifying{Evolution of the polarization conformity ratio as a function of the number of CSs circulating in the FP resonator. Experimental data are shown as diamonds, while the pink solid line corresponds to numerical simulations. Insets (\textit{i-ii}) display temporal traces of the CS sequence for 14 and 90 solitons circulating in the cavity, respectively.}}
\label{fig:conformity}
\end{figure}

Subsequently, FP resonators feature additional nonlocal nonlinear terms in Eqs.~(\ref{eq:LLEs}) that drive unique dynamics. In particular, the last extra four-wave mixing terms induce energy exchange between polarization components, with a strength roughly scaling with the average intra-cavity power. Remarkably, above a certain power threshold, this nonlinear coupling forces the circulating field into one of the two symmetry-broken states, giving rise to polarization conformity~\cite{hill_symmetry_2024}.
To demonstrate this phenomenon experimentally, we measured the probability for SB-CSs to converge to one of the two mirror-like asymmetric states shown in Fig.~\ref{fig:asym_sym}d as a function of their number in the resonator. For this purpose, the 20-GHz pulsed driving source was gated into 5-ns temporal bursts, whilst phase modulation was used to excite arbitrary numbers of CSs. We then measured on the fast oscilloscope the conformity ratio, \textit{i.e.} the probability for the whole CSs sequence to converge to the same asymmetric state. As shown in Fig.~\ref{fig:conformity}, for a moderate number of CSs, as exemplified by inset (\textit{i}), the CSs essentially behave independently, each one randomly selecting one of the two circular polarization states, yielding a near-zero probability of global polarization conformity. In contrast, as the number of CSs increases, a collective behavior gradually emerges and, for more than 80 CSs, essentially all of solitons end up in the same asymmetric state (conformity ratio $\approx 1$), as illustrated in inset (\textit{ii}). These results provide clear evidence that localized dissipative structures in FP resonators can exhibit collective polarization dynamics — a behavior fundamentally absent in ring geometries.

\begin{figure}[ht]
\centering
\includegraphics[width=\linewidth]{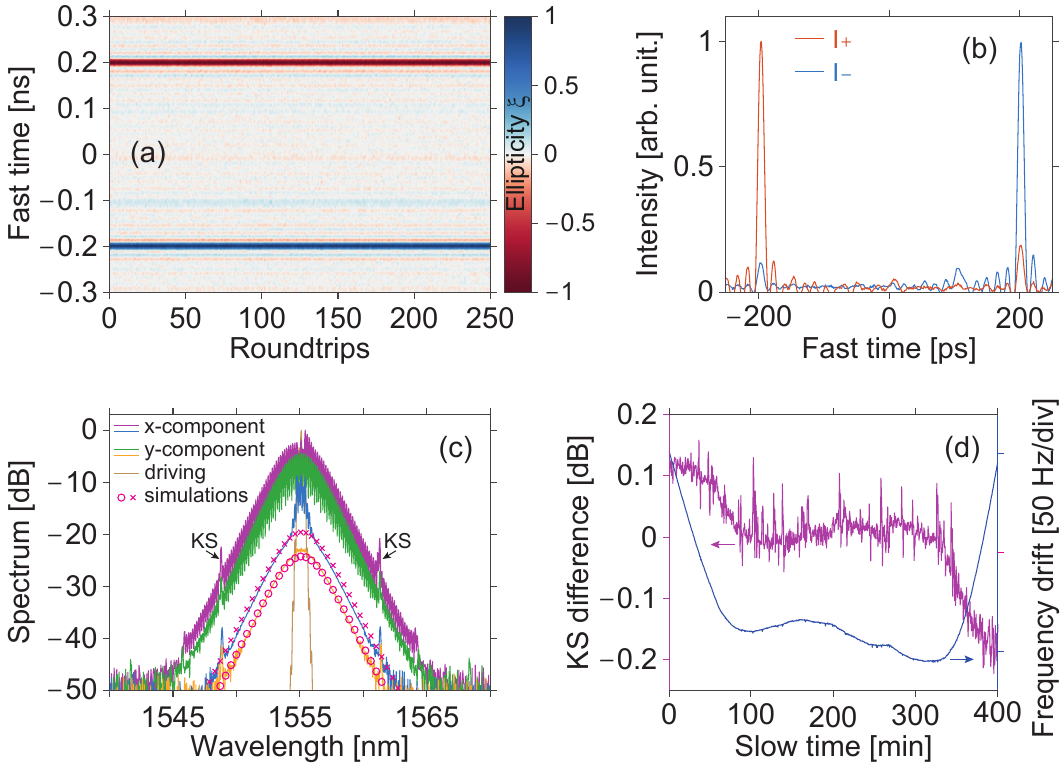}
\caption{\justifying{(a) Evolution of the polarization ellipticity $\xi$ over successive roundtrips with two SB-CSs circulating in the cavity. (b) Corresponding intensity profiles. (c) Optical spectra measured in the driving polarization basis: the blue and yellow curves correspond to the two CSs shown in Panels (a–b), while the purple and green curves show spectra for a sequence of 100 CSs. The brown curve represents the 20-GHz pulsed driving source, while the pink crosses and circles denote numerical simulations for a single CS. (d) Left axis in purple: power difference measured on the optical spectrum between the first Kelly sidebands (KS) of the CSs; right axis in dark blue: frequency correction applied on the pulsed driving repetition rate.}}
\label{fig:2CSs}
\end{figure}

%\subsection{Features of vectorial CSs.}
{To highlight in more detail the features of SB-CSs circulating in the FP, we then excited a pattern of two asymmetrical CSs. Figure~\ref{fig:2CSs}a shows the polarization ellipticity $\xi$ of the intracavity field over successive RTs. We can clearly observe two CSs of opposite handedness circulating endlessly in the cavity. Figure~\ref{fig:2CSs}b displays their intensity profiles, confirming the co-existence of two mirror-like solutions, both exhibiting strong polarization contrast. The corresponding optical spectra, projected onto the driving polarization basis, are shown in Fig.~\ref{fig:2CSs}c where the blue and yellow traces correspond to the two SB-CSs, while the brown curve indicates the 20-GHz pulsed driving source. These spectra confirm the presence of vectorial CSs in the resonator and exhibit a characteristic sech-shaped envelope with a 3-dB bandwidth of 0.35~THz --- consistent with numerical predictions, shown in pink, which yield a pulse width of 0.85~ps. Also shown are the output spectra obtained when a bunch of 100 SB-CSs are excited in the resonator, confirming the parametric generation of a 20-GHz repetition-rate, background-free CS train in the orthogonal (undriven) polarization component (green plot).

Finally, we would like to mention that long-term stability of these experiments was achieved by maintaining a perfect synchronization between the 20-GHz pulsed driving source and the cavity FSR. This was ensured by implementing a feedback loop that monitors the power asymmetry of the Kelly sidebands (KSs) in the CS spectrum (\textit{cf.} Fig.~\ref{fig:2CSs}c) and adjusts the initial RF clock to compensate for frequency drifts. As a result, we were able to maintain CS sequences in the cavity for several hours without any manual readjustment. Figure~\ref{fig:2CSs}d depicts a typical tracking of the power difference between the CS KSs (in purple) over nearly 7 hours as well as the corresponding correction applied on the RF clock (in dark blue). These data show that the maximum drift compensated with respect to the cavity FSR can reach 100~Hz, corresponding to a cumulative temporal desynchronization of 237~fs per roundtrip.}

%\section{Conclusion}
{In summary, we have experimentally and numerically demonstrated polarization symmetry-broken cavity solitons in a  coherently-driven passive Fabry–Pérot Kerr resonator. The FP cavity is made of a 5-m-long anomalous dispersive single-mode fiber, connected at both ends to highly reflective dielectric Bragg mirrors that are coated onto fiber connectors.
CSs are sustained in the resonator by means of a coherent pulse pumping scheme and exhibit spontaneous symmetry breaking of their polarization state induced by nonlinear cross-coupling interactions.
We observed the spontaneous transition of initially symmetric CSs into asymmetric vectorial states, with random selection of handedness. Most importantly, we experimentally confirmed a collective polarization conformity effect, whereby multiple SB-CSs circulating in the cavity are no longer independent entities but instead converge to the same asymmetric polarization state~\cite{hill_symmetry_2024}. This effect is analogous to polarization attraction in counter-propagating fiber systems, where birefringence reciprocity restricts the waves to right- or left-circular states~\cite{kaplan_light-induced_1983,pitois_polarization_1998,kozlov_nonlinear_2011}. This collective behavior, absent in ring geometries, highlights the unique nonlinear nonlocal coupling dynamics of Fabry–Pérot resonators and establishes them as a new platform for exploring vectorial and collective dissipative soliton physics.}

\begin{backmatter}
\bmsection{Funding} ANR Comby (ANR-25-CE24-4298-04), CNRS IRP WALL-IN, NSFC (62275097), Marsden Fund (18-UOA-310 and 23-UOA-053), the MQV Project TeQSiC, the German Federal Ministry of Research, Technology and Space, Quantum Systems, 13N17314, 13N17342, and the DFG project 541267874. AA  acknowledges the Max Planck School of Photonics and Florentina Gannott from TDSU Nanostructuring department.
\bmsection{Disclosures} The authors declare no conflicts of interest.
\bmsection{Data availability} The data that support the plots within this paper are available from the authors upon reasonable request.
\end{backmatter}

\bibliography{References}

@article{Leo2010,
	title = {Temporal cavity solitons in one-dimensional Kerr media as bits in an all-optical buffer},
	volume = {4},
	issn = {1749-4885},
	url = {http://www.nature.com/articles/nphoton.2010.120},
	doi = {10.1038/nphoton.2010.120},
	pages = {471--476},
	number = {7},
	journal = {Nature Photonics},
	author = {Leo, François and Coen, Stéphane and Kockaert, Pascal and Gorza, Simon-Pierre and Emplit, Philippe and Haelterman, Marc},
	year = {2010},
}

@article{Herr2013,
	title = {Temporal solitons in optical microresonators},
	volume = {8},
	issn = {1749-4885},
	url = {http://www.nature.com/doifinder/10.1038/nphoton.2013.343},
	doi = {10.1038/nphoton.2013.343},
	pages = {145--152},
	number = {2},
	journal = {Nature Photonics},
	author = {Herr, T. and Brasch, V. and Jost, J. D. and Wang, C. Y. and Kondratiev, N. M. and Gorodetsky, M. L. and Kippenberg, T. J.},
	year = {2013},
}

@article{Wabnitz1993,
	title = {Suppression of interactions in a phase-locked soliton optical memory},
	volume = {18},
	issn = {0146-9592},
	url = {http://www.ncbi.nlm.nih.gov/pubmed/19802213},
	doi = {10.1364/OL.18.000601},
	pages = {601--603},
	number = {8},
	journal = {Optics Letters},
	author = {Wabnitz, S},
	year = {1993},
	pmid = {19802213},
}

@article{kippenberg_dissipative_2018,
	title = {Dissipative {Kerr} solitons in optical microresonators},
	volume = {361},
	copyright = {Copyright © 2018, American Association for the Advancement of Science. http://www.sciencemag.org/about/science-licenses-journal-article-reuseThis is an article distributed under the terms of the Science Journals Default License.},
	issn = {0036-8075, 1095-9203},
	url = {http://science.sciencemag.org/content/361/6402/eaan8083},
	doi = {10.1126/science.aan8083},
	language = {en},
	number = {6402},
	journal = {Science},
	author = {Kippenberg, Tobias J. and Gaeta, Alexander L. and Lipson, Michal and Gorodetsky, Michael L.},
	month = aug,
	year = {2018},
	pmid = {30093576},
	pages = {eaan8083},
}

@article{oppo_theory_2024,
	title = {Theory and application of cavity solitons in photonic devices},
	volume = {382},
	issn = {1364-503X},
	url = {https://doi.org/10.1098/rsta.2023.0336},
	doi = {10.1098/rsta.2023.0336},
	number = {2287},
	urldate = {2026-02-23},
	journal = {Philosophical Transactions of the Royal Society A: Mathematical, Physical and Engineering Sciences},
	author = {Oppo, Gian-Luca and Firth, William J.},
	month = dec,
	year = {2024},
	pages = {20230336},
}

@article{pasquazi_micro-combs_2018,
	series = {Micro-combs: {A} novel generation of optical sources},
	title = {Micro-combs: {A} novel generation of optical sources},
	volume = {729},
	issn = {0370-1573},
	shorttitle = {Micro-combs},
	url = {http://www.sciencedirect.com/science/article/pii/S0370157317303253},
	doi = {10.1016/j.physrep.2017.08.004},
	urldate = {2018-02-14},
	journal = {Physics Reports},
	author = {Pasquazi, Alessia and Peccianti, Marco and Razzari, Luca and Moss, David J. and Coen, Stéphane and Erkintalo, Miro and Chembo, Yanne K. and Hansson, Tobias and Wabnitz, Stefan and Del’Haye, Pascal and Xue, Xiaoxiao and Weiner, Andrew M. and Morandotti, Roberto},
	month = jan,
	year = {2018},
	pages = {1--81},
}

@article{Obrzud2017,
	title = {Temporal solitons in microresonators driven by optical pulses},
	volume = {11},
	issn = {17494893},
	url = {http://dx.doi.org/10.1038/nphoton.2017.140},
	doi = {10.1038/nphoton.2017.140},
	pages = {600--607},
	number = {9},
	journal = {Nature Photonics},
	author = {Obrzud, Ewelina and Lecomte, Steve and Herr, Tobias},
	year = {2017},
	pmid = {23099347},
}

@article{Englebert2020,
	title = {Temporal solitons in a coherently driven active resonator},
	volume = {15},
	issn = {1749-4885},
	url = {http://www.nature.com/articles/s41566-021-00807-w},
	doi = {10.1038/s41566-021-00807-w},
	pages = {536--541},
	number = {7},
	journal = {Nature Photonics},
	author = {Englebert, Nicolas and Mas Arabí, Carlos and Parra-Rivas, Pedro and Gorza, Simon-Pierre and Leo, François},
	year = {2021},
}

@article{pedaci_positioning_2006,
	title = {Positioning cavity solitons with a phase mask},
	volume = {89},
	issn = {0003-6951, 1077-3118},
	url = {http://scitation.aip.org/content/aip/journal/apl/89/22/10.1063/1.2388867},
	doi = {10.1063/1.2388867},
	number = {22},
	journal = {Applied Physics Letters},
	author = {Pedaci, F. and Genevet, P. and Barland, S. and Giudici, M. and Tredicce, J. R.},
	month = nov,
	year = {2006},
	pages = {221111/1--3},
}

@article{lilienfein_temporal_2019,
	title = {Temporal solitons in free-space femtosecond enhancement cavities},
	volume = {13},
	issn = {1749-4893},
	url = {https://www.nature.com/articles/s41566-018-0341-y},
	doi = {10.1038/s41566-018-0341-y},
	language = {English},
	number = {3},
	journal = {Nat. Photon.},
	author = {Lilienfein, N. and Hofer, C. and Högner, M. and Saule, T. and Trubetskov, M. and Pervak, V. and Fill, E. and Riek, C. and Leitenstorfer, A. and Limpert, J. and Krausz, F. and Pupeza, I.},
	month = mar,
	year = {2019},
	pages = {214--218},
}

@article{Coen_fat_2024,
	title = {Nonlinear topological symmetry protection in a dissipative system},
	volume = {15},
	rights = {2024 The Author(s)},
	issn = {2041-1723},
	url = {https://www.nature.com/articles/s41467-023-44640-x},
	doi = {10.1038/s41467-023-44640-x},
	pages = {1398},
	number = {1},
	journal = {Nature Communications},
	shortjournal = {Nature Commun.},
	author = {Coen, Stéphane and Garbin, Bruno and Xu, Gang and Quinn, Liam and Goldman, Nathan and Oppo, Gian-Luca and Erkintalo, Miro and Murdoch, Stuart G. and Fatome, Julien},
	year = {2024},
	langid = {english},
}

@article{xu_spontaneous_2021,
	title = {Spontaneous symmetry breaking of dissipative optical solitons in a two-component Kerr resonator},
	volume = {12},
	issn = {2041-1723},
	url = {http://www.nature.com/articles/s41467-021-24251-0},
	doi = {10.1038/s41467-021-24251-0},
	pages = {4023},
	number = {1},
	journal = {Nature Communications},
	author = {Xu, Gang and Nielsen, Alexander U. and Garbin, Bruno and Hill, Lewis and Oppo, Gian-Luca and Fatome, Julien and Murdoch, Stuart G. and Coen, Stéphane and Erkintalo, Miro},
	year = {2021},
	pmid = {34188030},
}

@article{Hill2020,
	title = {Breathing dynamics of symmetry-broken temporal cavity solitons in Kerr ring resonators},
	volume = {47},
	issn = {0146-9592},
	url = {https://opg.optica.org/abstract.cfm?URI=ol-47-6-1486},
	doi = {10.1364/OL.449679},
	pages = {1486--1489},
	number = {6},
	journal = {Optics Letters},
	author = {Xu, Gang and Hill, Lewis and Fatome, Julien and Oppo, Gian-Luca and Erkintalo, Miro and Murdoch, Stuart G. and Coen, Stéphane},
	year = {2022},
}

@article{lucas_polarization_2025,
	title = {Polarization {Faticons}: {Chiral} {Localized} {Structures} in {Self}-{Defocusing} {Kerr} {Resonators}},
	volume = {135},
	shorttitle = {Polarization {Faticons}},
	url = {https://link.aps.org/doi/10.1103/ljbj-tz7g},
	doi = {10.1103/ljbj-tz7g},
	number = {6},
	journal = {Physical Review Letters},
	author = {Lucas, Erwan and Xu, Gang and Wang, Pengxiang and Oppo, Gian-Luca and Hill, Lewis and Del’Haye, Pascal and Kibler, Bertrand and Xu, Yiqing and Murdoch, Stuart G. and Erkintalo, Miro and Coen, Stéphane and Fatome, Julien},
	month = aug,
	year = {2025},
	pages = {063803},
}

@article{moroney_kerr_2022,
	title = {A Kerr polarization controller},
	volume = {13},
	rights = {2022 The Author(s)},
	issn = {2041-1723},
	url = {https://www.nature.com/articles/s41467-021-27933-x},
	doi = {10.1038/s41467-021-27933-x},
	pages = {398},
	number = {1},
	journal = {Nature Communications},
	shortjournal = {Nat Commun},
	author = {Moroney, N. and Del Bino, L. and Zhang, S. and Woodley, M. T. M. and Hill, L. and Wildi, T. and Wittwer, V. J. and Südmeyer, T. and Oppo, G.-L. and Vanner, M. R. and Brasch, V. and Herr, T. and Del’Haye, P.},
	urldate = {2023-04-28},
	year = {2022},
	langid = {english},
}

@article{yang_polarization_2025,
	title = {Polarization {Symmetry} {Breaking} of {GHz} {Dissipative} {Solitons}},
	volume = {134},
	url = {https://link.aps.org/doi/10.1103/x3sy-gf7l},
	doi = {10.1103/x3sy-gf7l},
	number = {21},
	journal = {Physical Review Letters},
	author = {Yang, Yang and Chen, Xuewen and Lin, Wei and Hu, Xu and Xu, Haijiao and Ma, Yuncong and Liang, Zhaoheng and Ling, Lin and Xiong, Zhijin and Guo, Yuankai and Liu, Tao and Wei, Xiaoming and Yang, Zhongmin},
	year = {2025},
	pages = {213803},
}

@article{jia_photonic_2020,
	title = {Photonic {Flywheel} in a {Monolithic} {Fiber} {Resonator}},
	volume = {125},
	issn = {0031-9007, 1079-7114},
	url = {https://link.aps.org/doi/10.1103/PhysRevLett.125.143902},
	doi = {10.1103/PhysRevLett.125.143902},
	language = {en},
	number = {14},
	journal = {Physical Review Letters},
	author = {Jia, Kunpeng and Wang, Xiaohan and Kwon, Dohyeon and Wang, Jiarong and Tsao, Eugene and Liu, Huaying and Ni, Xin and Guo, Jian and Yang, Mufan and Jiang, Xiaoshun and Kim, Jungwon and Zhu, Shi-ning and Xie, Zhenda and Huang, Shu-Wei},
	month = oct,
	year = {2020},
	pages = {143902},
}

@article{musgrave_microcombs_2023,
	title = {Microcombs in fiber {Fabry}–{Pérot} cavities},
	volume = {8},
	issn = {2378-0967},
	url = {https://doi.org/10.1063/5.0177134},
	doi = {10.1063/5.0177134},
	number = {12},
	journal = {APL Photonics},
	author = {Musgrave, Jonathan and Huang, Shu-Wei and Nie, Mingming},
	year = {2023},
	pages = {121101},
}

@article{li_ultrashort_2024,
	title = {Ultrashort dissipative {Raman} solitons in {Kerr} resonators driven with phase-coherent optical pulses},
	volume = {18},
	copyright = {2023 The Author(s), under exclusive licence to Springer Nature Limited},
	issn = {1749-4893},
	url = {https://www.nature.com/articles/s41566-023-01303-z},
	doi = {10.1038/s41566-023-01303-z},
	language = {en},
	number = {1},
	journal = {Nature Photonics},
	author = {Li, Zongda and Xu, Yiqing and Shamailov, Sophie and Wen, Xiaoxiao and Wang, Wenlong and Wei, Xiaoming and Yang, Zhongmin and Coen, Stéphane and Murdoch, Stuart G. and Erkintalo, Miro},
	year = {2024},
	pages = {46--53},
}

@article{bunel_28_2024,
	title = {28 {THz} soliton frequency comb in a continuous-wave pumped fiber {Fabry}–{Pérot} resonator},
	volume = {9},
	issn = {2378-0967},
	url = {https://doi.org/10.1063/5.0176533},
	doi = {10.1063/5.0176533},
	number = {1},
	journal = {APL Photonics},
	author = {Bunel, T. and Conforti, M. and Ziani, Z. and Lumeau, J. and Moreau, A. and Fernandez, A. and Llopis, O. and Bourcier, G. and Mussot, A.},
	year = {2024},
	pages = {010804},
}

@article{cole_theory_2018,
	title = {Theory of {Kerr} frequency combs in {Fabry}-{Perot} resonators},
	volume = {98},
	url = {https://link.aps.org/doi/10.1103/PhysRevA.98.013831},
	doi = {10.1103/PhysRevA.98.013831},
	number = {1},
	journal = {Physical Review A},
	author = {Cole, Daniel C. and Gatti, Alessandra and Papp, Scott B. and Prati, Franco and Lugiato, Luigi},
	year = {2018},
	pages = {013831},
}

@article{hill_symmetry_2024,
	title = {Symmetry broken vectorial {Kerr} frequency combs from {Fabry}-{Pérot} resonators},
	volume = {7},
	copyright = {2024 The Author(s)},
	issn = {2399-3650},
	url = {https://www.nature.com/articles/s42005-024-01566-0},
	doi = {10.1038/s42005-024-01566-0},
	language = {en},
	number = {1},
	journal = {Communications Physics},
	author = {Hill, Lewis and Hirmer, Eva-Maria and Campbell, Graeme and Bi, Toby and Ghosh, Alekhya and Del’Haye, Pascal and Oppo, Gian-Luca},
	year = {2024},
	pages = {1--9},
}

@article{fatome_20ghz_2006,
	title = {20-{GHz}-to-1-{THz} {Repetition} {Rate} {Pulse} {Sources} {Based} on {Multiple} {Four}-{Wave} {Mixing} in {Optical} {Fibers}},
	volume = {42},
	issn = {1558-1713},
	url = {https://ieeexplore.ieee.org/document/1703695},
	doi = {10.1109/JQE.2006.881826},
	number = {10},
	journal = {IEEE Journal of Quantum Electronics},
	author = {Fatome, J. and Pitois, S. and Millot, G.},
	year = {2006},
	pages = {1038--1046},
}

@article{jang_writing_2015,
	title = {Writing and erasing of temporal cavity solitons by direct phase modulation of the cavity driving field},
	volume = {40},
	copyright = {© 2015 Optical Society of America},
	issn = {1539-4794},
	url = {http://www.osapublishing.org/abstract.cfm?uri=ol-40-20-4755},
	doi = {10.1364/OL.40.004755},
	number = {20},
	urldate = {2015-10-13},
	journal = {Optics Letters},
	author = {Jang, Jae K. and Erkintalo, Miro and Murdoch, Stuart G. and Coen, Stéphane},
	month = oct,
	year = {2015},
	pages = {4755--4758},
}

@article{kaplan_light-induced_1983,
	title = {Light-induced nonreciprocity, field invariants, and nonlinear eigenpolarizations},
	volume = {8},
	copyright = {\&\#169; 1983 Optical Society of America},
	issn = {1539-4794},
	url = {https://www.osapublishing.org/ol/abstract.cfm?uri=ol-8-11-560},
	doi = {10.1364/OL.8.000560},
	number = {11},
	urldate = {2018-04-11},
	journal = {Optics Letters},
	author = {Kaplan, A. E.},
	month = nov,
	year = {1983},
	pages = {560--562},
}

@article{kozlov_nonlinear_2011,
	title = {Nonlinear repolarization dynamics in optical fibers: transient polarization attraction},
	volume = {28},
	copyright = {© 2011 Optical Society of America},
	issn = {1520-8540},
	shorttitle = {Nonlinear repolarization dynamics in optical fibers},
	url = {https://opg.optica.org/josab/abstract.cfm?uri=josab-28-8-1782},
	doi = {10.1364/JOSAB.28.001782},
	language = {EN},
	number = {8},
	urldate = {2026-02-24},
	journal = {JOSA B},
	publisher = {Optica Publishing Group},
	author = {Kozlov, Victor V. and Fatome, Julien and Morin, Philippe and Pitois, Stephane and Millot, Guy and Wabnitz, Stefan},
	month = aug,
	year = {2011},
	pages = {1782--1791},
}

@article{pitois_polarization_1998,
	title = {Polarization domain wall solitons with counterpropagating laser beams},
	volume = {81},
	url = {https://journals.aps.org/prl/abstract/10.1103/PhysRevLett.81.1409},
	doi = {10.1103/PhysRevLett.81.1409},
	number = {7},
	journal = {Physical Review Letters},
	author = {Pitois, S. and Millot, Guy and Wabnitz, Stefan},
	year = {1998},
	pages = {1409--1412},
}

@article{hill2020effects,
  title={Effects of self-and cross-phase modulation on the spontaneous symmetry breaking of light in ring resonators},
  author={Hill, Lewis and Oppo, Gian-Luca and Woodley, Michael TM and Del'Haye, Pascal},
  journal={Physical Review A},
  volume={101},
  number={1},
  pages={013823},
  year={2020},
  publisher={APS}
}

@article{campbell2024frequency,
  title={Frequency comb enhancement via the self-crystallization of vectorial cavity solitons},
  author={Campbell, Graeme N and Hill, Lewis and Del’Haye, Pascal and Oppo, Gian-Luca},
  journal={Optics Express},
  volume={32},
  number={21},
  pages={37691--37702},
  year={2024},
  publisher={Optica Publishing Group}
}
% Full bibliography added automatically for Optics Letters submissions; the following line will simply be ignored if submitting to other journals.
% Note that this extra page will not count against page length
\bibliographyfullrefs{References}

\end{document}